\documentclass[runningheads]{llncs}
\usepackage{graphicx}
\usepackage{array} 
\usepackage{amsmath}
\usepackage{amssymb}
\usepackage{algorithm}
\usepackage{algorithmic}
\usepackage{xspace}
\usepackage{url}
\usepackage{booktabs}
\usepackage{subfig}
\usepackage{cite}
\usepackage{wrapfig}
\newcommand{\name}{\textsc{Chronosymbolic Learning}\xspace}

\begin{document}
\title{Chronosymbolic Learning: \\Efficient CHC Solving with\\Symbolic Reasoning and Inductive Learning}
\titlerunning{Chronosymbolic Learning}
%
\author{Ziyan Luo\inst{1,2} \and
Xujie Si\inst{1,3} }
%
\authorrunning{Ziyan Luo \and Xujie Si}
%
\institute{Mila \and McGill University \and University of Toronto \\
\email{ziyan.luo@mail.mcgill.ca, six@cs.toronto.edu}}
\maketitle

\begin{abstract}
Solving Constrained Horn Clauses (CHCs) is a fundamental challenge behind a wide range of verification and analysis tasks. 
To enhance CHC solving without the laborious task of manual heuristic creation and tuning, data-driven approaches demonstrate significant potential by extracting crucial patterns from a small set of data points.
However, at present, symbolic methods generally surpass data-driven solvers in performance.
In this work, we develop a simple but effective framework, \name, which unifies symbolic information and numerical data to solve a CHC system efficiently.
We also present a simple instance\footnote{The artifact is available on this link: \url{https://github.com/Chronosymbolic/Chronosymbolic-Learning}} of \name with a data-driven learner and a BMC-styled reasoner\footnote{BMC represents Bounded Model Checking \cite{bmc}. }. Despite its relative simplicity, experimental results show the efficacy and robustness of our tool. It outperforms state-of-the-art CHC solvers on a test suite of 288 arithmetic benchmarks, including some instances with non-linear arithmetic.

\end{abstract}

\section{Introduction}
Constrained Horn Clauses (CHCs),  a fragment of First Order Logic (FOL), naturally capture the discovery and verification of inductive invariants~\cite{bjorner2015horn}.
CHCs serve as a general format for program safety verification\footnote{See Appendix \ref{CHC_program_verfication} for details.}. They are widely used in software verification frameworks including 
C/C++, Java, Rust, Solidity, and Android verification frameworks \cite{gurfinkel2015seahorn,kahsai2016jayhorn,rusthorn}, modular verification of distributed and parameterized systems~\cite{pldi12:GrebenshchikovLPR,Gurfinkel2016fse}, type inference~\cite{Soltype:popl12}, and many others~\cite{gurfinkel2022program}.
Given the importance of these applications, building an efficient CHC solver holds great significance. 
Nevertheless, the undecidability of CHC solving necessitates tailored adjustments or designs for specific instances, demanding substantial effort and expertise. 

\begin{figure}
\centering
\vspace{-0.4cm}
\includegraphics[width=0.8\textwidth]{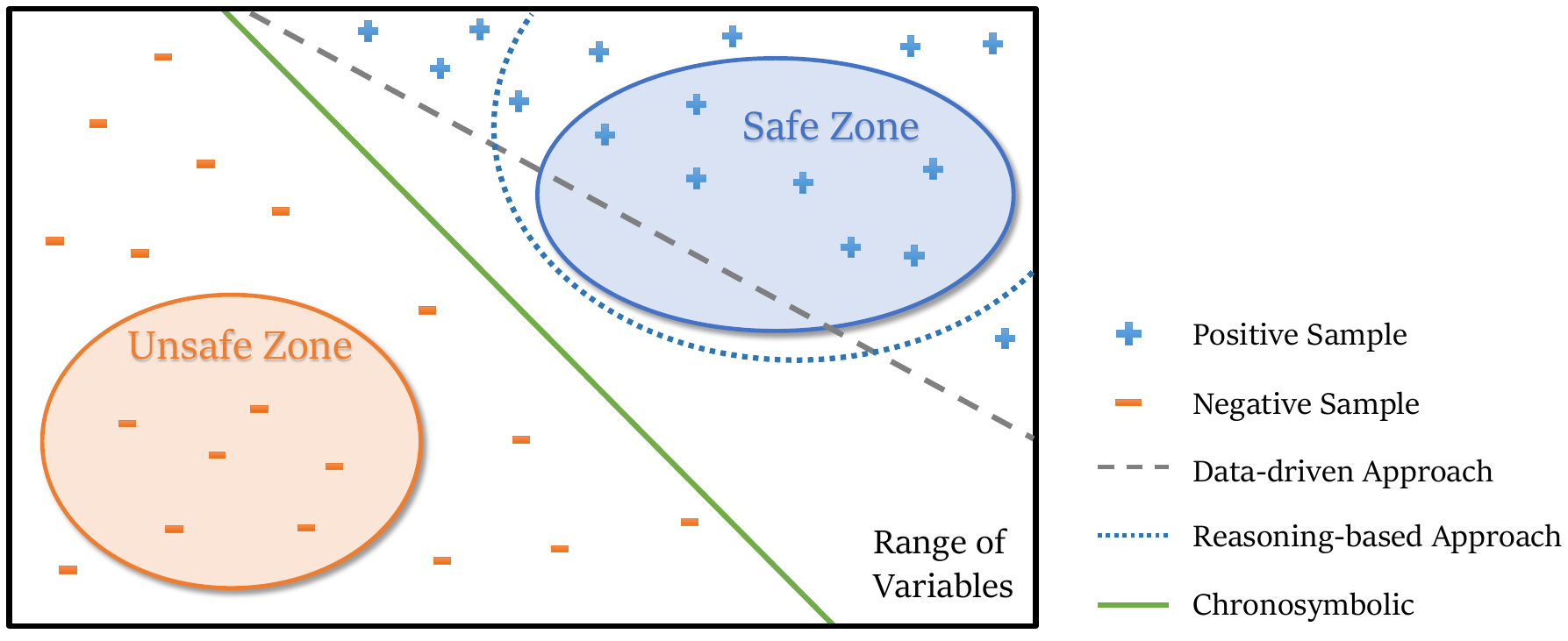}
\caption{Overview of different approaches through the lens of learning from positive and negative samples. 
} \label{3solvers}
\vspace{-0.4cm}
\end{figure}

Remarkable progress in automating CHC solving has been achieved in recent years. Existing approaches primarily fall into two categories: symbolic-reasoning-based approaches and data-driven induction-based approaches. 
The former centers on designing novel symbolic reasoning techniques, such as abstraction refinement~\cite{cegar}, interpolation~\cite{mcmillan2003interpolation}, property-directed reachability~\cite{bradley2011sat,pdr:fmcad11}, model-based projections~\cite{komuravelli2013automatic}, and other techniques. 
While the latter focuses on reducing CHC solving into a machine learning (ML) problem and then employing proper ML models, such as Boolean functions~\cite{padhi:pldi16}, decision trees (DTs)~\cite{garg2016learning}, support vector machines (SVMs)~\cite{zhu2018data}, and deep learning models~\cite{si2020code2inv}.
While data-driven approaches show great promise towards improving CHC solving without the painstaking manual heuristic tuning, data-driven CHC solvers still fall way behind symbolic reasoning-based CHC solvers~\cite{vediramana2020global}. 
Fig.~\ref{3solvers} illustrates key differences between these two categories.
Symbolic approaches usually maintain two \textit{zones} approximating safe and unsafe ``states'' of a system represented by CHCs, meticulously updated with soundness guarantees. 
Data-driven approaches abstract symbolic constraints away by sampling positive and negative data points. 
The sampling process usually requires some form of \textit{evaluation} of the given constraints, thus, sampled data points are the outcome of complicated interactions among multiple constraints, making data-driven approaches excel at capturing useful global properties.
However, as illustrated by the dashed grey line in Fig.~\ref{3solvers}, 
the primary drawback of data-driven approaches is that the classifier learned from samples may overlook safe regions that symbolic reasoning can easily identify. 
Conversely, symbolic approaches are good at precisely analyzing local constraints, delivering superior performance but potentially getting stuck in a local region. 
Additionally, 
they struggle to identify essential patterns from data samples, limiting their flexibility in addressing these problems.


The motivation for our framework, \name, is to devise a learning framework integrating the strengths of both symbolic and data-driven methods.
Similar to data-driven approaches, a \textit{learner} in \name derives classifiers from sampled data points, a process attainable by standard machine learning approaches. 
However, rather than utilizing these learned classifiers directly as hypotheses, we augment them with symbolic information summarized by a \textit{reasoner}.
Such symbolic information can be viewed as a region of data points \textit{concisely} expressed in a symbolic form, whereas the learned classifiers serve as symbolic representations of \textit{generalized} regions.
Instead of combining these two mechanically or parallelly, our method makes the reasoner and learner mutually benefit from each other, as exemplified by the green line of Fig.~\ref{3solvers}.  
It also lays the groundwork for seamlessly synergizing the power of data learning with automated reasoning, offering a paradigm that guides future efforts to incorporate cutting-edge ML techniques in CHC solving.

\paragraph{Main Contributions.}
Our contribution is mainly three-fold.
First and foremost, we propose a novel framework, \name. 
As the name suggests\footnote{The name \textsc{\underline{Ch}ronosymboli\underline{c} Learning} is a blend of the terms  CHC, number (represented by ``no.''), symbolic and synchronous.}, 
our goal is to solve CHCs with both numerical data samples and symbolic information synchronously and synergistically. 
We provide formulations in order to realize this desideratum, establishing the groundwork for the application of advanced techniques in symbolic reasoning and machine learning.
Secondly, we build a simple yet potent instance of \name in Python, the standard programming language in the ML community, to substantiate our claims. It comprises a data-driven \textit{learner} interacting with a BMC-styled \textit{reasoner}, alongside a verification oracle as a \textit{teacher}.
We also provide a discussion on alternative design choices.
Lastly, the main experimental results demonstrate the effectiveness and robustness of our tool. It outperforms several state-of-the-art CHC solvers on a test suite of 288 benchmarks, including some instances with non-linear integer arithmetic.
\section{Preliminaries}
\subsection{Constrained Horn Clauses}
\label{chc}
We discuss standard \textit{First Order Logic} (FOL) formula modulo theory $\mathcal{T}$, with a signature $\Sigma$ composed by constant symbols $\mathcal{A}$ (e.g., True $\top$ and False $\bot$), function symbols $\mathcal{F}$ (e.g., +, -, \texttt{mod}) and predicate symbols $\mathcal{P}$. A \textit{Constrained Horn Clause} (CHC) $\mathcal{C}$ is a FOL formula modulo theory $\mathcal{T}$ in the following form:
\begin{equation}
    \forall \mathcal{X} \cdot \phi  \land p_1(T_1) \land \cdots \land p_k(T_k) \rightarrow h(T), \ k \geq 0,
\end{equation}
where $\mathcal{X}$ stands for all variables in $T_i$ and $T$, $\phi$ represents a fixed constraint over $\mathcal{F},\mathcal{A} \text{ and } \mathcal{X}$ w.r.t. some background theory $\mathcal{T}$; 
$p_i, h$ are uninterpreted predicate symbols\footnote{``Predicates'' for short.}, and $p_i(T_i)=p_i(t_{i,1},\cdots,t_{i,n})$ is an application of n-ary predicate symbol $p_i$ with first-order terms $t_{i,j}$ built from $\mathcal{F},\mathcal{A} \text{ and } \mathcal{X}$;
$h(T)$ could be either $\mathcal{P}$-free (i.e., no predicate symbol appears in $h$) or akin to the definition of $p_i(T_i)$. 

In this work, we explain our framework using the background theory of \textit{Integer Arithmetic} (IA) if it is not specified.
For simplification, in our work, the universal quantifier ``$\forall$'' is omitted when the context is clear; and all terms $t_{i,j}$ represent single variables (e.g., $t_{1,1}=x$, $t_{1,2}=y$).

The structure of a CHC can be viewed as a split based on the implication symbol. The left hand side $\phi  \wedge p_1(T_1) \wedge \cdots \wedge p_k(T_k)$ is called \textit{body} and the right hand side $h(T)$ is called \textit{head}.
If the body of a CHC contains zero or one predicate symbol, the CHC is called \textit{linear}. Otherwise, it is called \textit{non-linear}.

\begin{example}
\label{exp}
A simple CHC system $\mathcal{H}_0$\footnote{This is the benchmark \texttt{nonlin\_mult\_2.smt2} in our suite of test benchmarks.} consisting of a fact $\mathcal{C}_0$, a non-fact rule $\mathcal{C}_1$ and a query $\mathcal{C}_2$ is as follows:

\vspace{-0.55cm}
\small
\begin{align*}
&\mathcal{C}_0:\neg (a \leq 0) \ \land \  b \leq a \ \land \ c=0 \ \land \ e=0 \ \land \ d=0 \rightarrow inv(a,b,c,d,e)\\
&\mathcal{C}_1:c_1=1+c\land d_1=d+a\land e_1=e+b\land inv(a,b,c,d,e) \rightarrow inv(a,b,c_1,d_1,e_1)\\
&\mathcal{C}_2:\neg (e \geq a \cdot c) \ \land \  inv(a,b,c,d,e) \rightarrow \bot
\end{align*}
\end{example}

In most cases, the terminology ``CHC solving'' refers to solving the \textit{satisfiability} (SAT) problem for a \textit{CHC system} that is a set of CHCs containing at least a query and a rule. A \textit{query} $\mathcal{C}_q$ refers to a CHC that has a $\mathcal{P}$-free head, otherwise, it is called a \textit{rule} $\mathcal{C}_r$. Particularly, a \textit{fact} $\mathcal{C}_f$ refers to a rule that has a $\mathcal{P}$-free body. 
For instance, in Example \ref{exp}, $\mathcal{C}_0$ is a fact and a rule,  $\mathcal{C}_1$ is a rule, and $\mathcal{C}_2$ is a query.
We define the clauses that both body and head are $\mathcal{P}$-free as \textit{trivial} clauses because they can be simply reduced to $\top$ or $\bot$. 

Then, we formally define the satisfiability of a CHC system $\mathcal{H}$:



\begin{definition}[Satisfiability of a CHC system]
$\mathcal{H}$ is satisfiable \textnormal{(SAT)} iff there exists an interpretation of predicates $\mathcal{I}^*$ such that $\forall \mathcal{C} \in \mathcal{H}$, $\mathcal{I}^*\left[\mathcal{C}\right]$ is $\top$. Otherwise, we say $\mathcal{H}$ is unsatisfiable \textnormal{(UNSAT)}.
\end{definition}

\begin{definition}[Solution of a CHC system]
A solution of $\mathcal{H}$ consists of its satisfiability and a proof of its satisfiability.
\end{definition}
If $\mathcal{H}$ is SAT, the proof is the corresponding interpretation $\mathcal{I}^*$, and $\mathcal{I}^*[p]$ is called a \textit{solution interpretation} of the predicate $p$. If  $\mathcal{H}$ is UNSAT, the proof is a \textit{refutation} $\mathcal{R}$ demonstrating the nonexistence of a solution interpretation \cite{bjorner2013solving}.
Therefore, the problem addressed in this work is to \textit{determine a solution for a given CHC system}.

\subsection{CHC Solving as a Symbolic Classification Problem}
\label{mltool}
To harness data samples, we cast the CHC solving problem as a symbolic classification task by utilizing the samples to produce hypotheses. 
For clarity, we formally define the term \textit{hypothesis} of a CHC system $\mathcal{H}$:
\begin{definition}[Hypothesis]
\label{hyp}
 A hypothesis of $\mathcal{H}$ is an interpretation $\Tilde{\mathcal{I}}$ that is a possible solution interpretation of $\mathcal{H}$. 
\end{definition} 
To determine a hypothesis for a specific predicate $p$ denoted as $\Tilde{\mathcal{I}}\left[p\right]$, \cite{zhu2018data} introduces a lightweight machine learning toolchain that combines DT learning with the results of SVMs. 
In this data-driven pipeline, positive and negative samples are iteratively classified by SVMs until all samples are correctly categorized. 
The resulting hyperplanes of SVMs, $f(\boldsymbol{x})=w^T \boldsymbol{x}_i+b$, serve as \textit{attributes} for the DT.
The DT then selects an attribute $f(\boldsymbol{x}_0)$ and a corresponding threshold $c$ in the form of $f(\boldsymbol{x}_0)\leq c$ to create a new node in the way of maximizing the \textit{information gain} $\gamma$.
Adjusting the threshold $c$ and pruning the over-complicated attribute combinations enhances generalizability and mitigates overfitting risks. 
This process forms a DT, effectively segregating all samples into positive and negative categories. 
The DT generates a learned formula\footnote{The fundamentals of DT, SVM, and the formula generation is specified in Appendix~\ref{app:extpre}.}, denoted as $\mathcal{L}_p$, which can be arbitrary linear inequalities and their combinations in \textit{disjunctive normal form} (DNF). 
In our framework, this formula is termed a \textit{partition}: 
a hypothesis proposed only by the learner.

\section{Chronosymbolic Learning}
As shown in Fig. \ref{3solvers}, pure data-driven and reasoning-based approaches exhibit distinct limitations.
The former is completely agnostic to the inherent symbolic information within the CHC system, and relies solely on data samples for hypothesis generation.
The latter struggles to \textit{induce} global patterns of solution interpretations, and faces challenges in effectively integrating data samples into its reasoning process.
To address the limitations, we propose \name, a modular framework designed to amalgamate the strengths of both approaches and harness the full potential of symbolic information and numerical data concurrently.

\vspace{-0.2cm}
\begin{figure}
\centering
\includegraphics[width=0.7\textwidth]{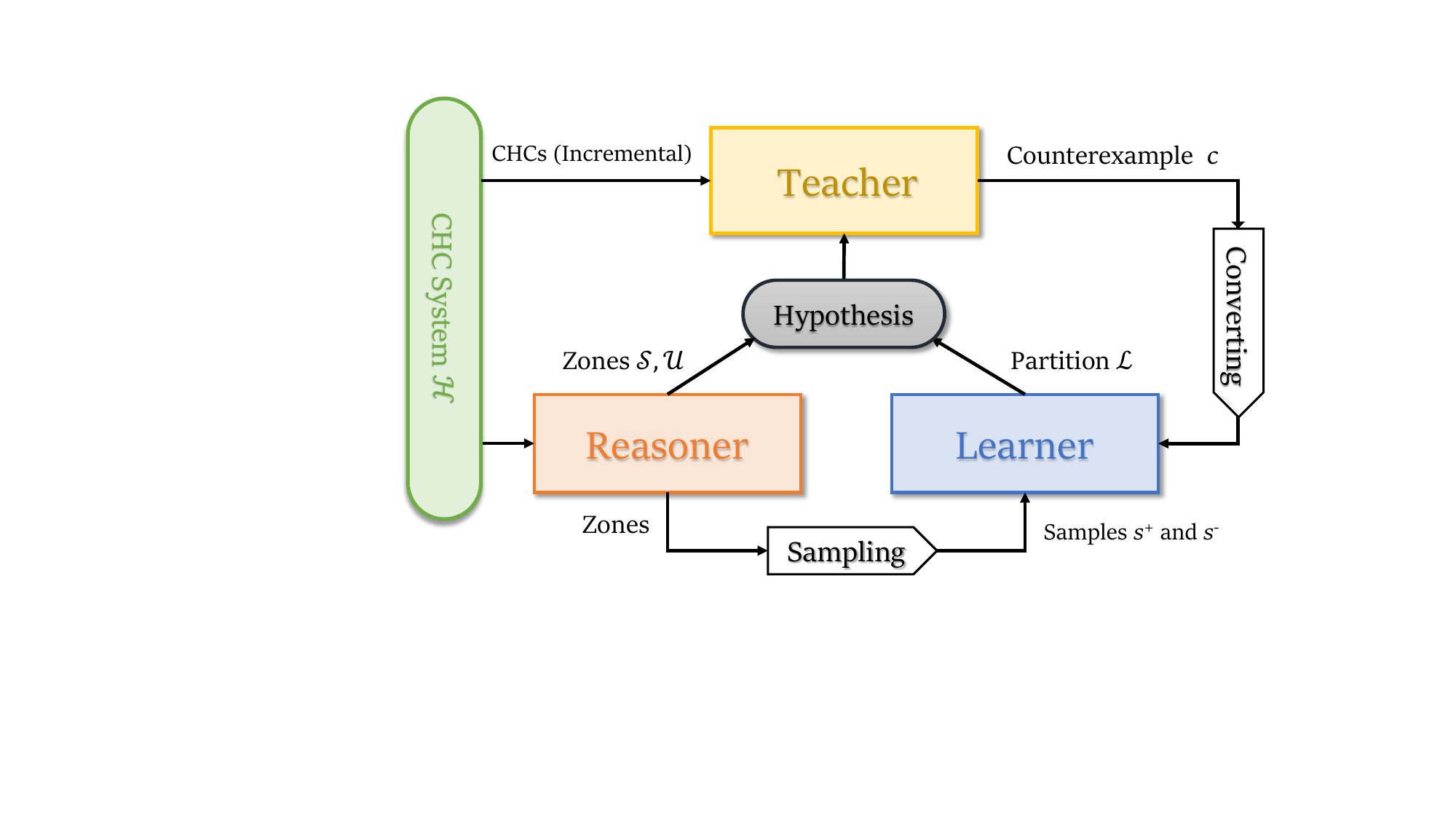}
\caption{
The architecture of \name. 
} \label{arc}
\end{figure}

\vspace{-1cm}
\subsection{Architecture of Chronosymbolic Learning}
\name, depicted in Fig. \ref{arc}, extends the paradigm of ``\textit{teacher\footnote{Also refers to a verification oracle, such as Microsoft Z3.} and learner}''  \cite{garg2014ice}, which solves a given CHC system by guessing and checking \cite{houdini, sharma2013data}. 
In this paradigm, the teacher and learner engage in iterative communication. 
During each iteration, the learner formulates a \textit{hypothesis} (refer to Definition \ref{hyp}) of an interpretation called \textit{partition}. The teacher then verifies the hypothesis and provides instant feedback regarding its correctness. If incorrect, the teacher supplies a \textit{counterexample} (see Definition \ref{cex}) elucidating the reason for its inaccuracy.

In addition to this paradigm, in \name, the learning procedure is enhanced by a \textit{reasoner}.
The reasoner maintains a \textit{safe zone} and an \textit{unsafe zone}, representing symbolic equivalents of \textit{positive and negative samples} respectively. These zones offer three main advantages: 1) They can be integrated into the learner's proposed hypothesis to enhance it; 2) They provide the learner with additional samples; 3) They significantly simplify the UNSAT checking of the CHC system\footnote{In program verification parlance, it refers to the unsafe check.}. 
We will provide an instance of \name,  detailing the functionality and intercommunication of its modules as in Fig. \ref{arc}. The learner we specify in Section \ref{sec:learner} and reasoner in Section \ref{sec:reasoner} can be replaced by other algorithms capable of generating partitions and zones, as discussed in Section \ref{sec:discussion}. 
See Section \ref{algor} for the overall algorithm.

\subsection{Samples and Zones}
To introduce \name, we first conceptualize samples and zones. We shed light on obtaining the samples and zones in Section \ref{sec:design_of_lr}.

Positive samples and negative samples in our framework are defined as generalizations of reachable program states from facts and queries, and implication samples are defined as borrowed from the concept in \cite{garg2014ice}.

\begin{definition}[Positive Sample]
\label{possamp}
A data point $s^+$ is a positive sample of predicate $p$ in $\mathcal{H}$ iff $p(s^+)=\top$ must hold to make all rules in $\mathcal{H}$ SAT.
\end{definition}

\begin{definition}[Negative Sample]
\label{negsamp}
A data point $s^-$ is a negative sample of $p$ in $\mathcal{H}$ iff $p(s^-)=\bot$ must hold to make all non-fact rules and queries in $\mathcal{H}$ SAT.
\end{definition}

\begin{definition}[Implication Sample]
\label{def:impl}
An implication sample $s^{\rightarrow}$
of body predicates $(p_1,...,p_n)$ and head predicate $h$ in $\mathcal{H}$ is an $(n+1)$-tuple of data points $\left(s^\rightarrow_{1},\cdots,s^\rightarrow_{n},s^{\rightarrow}_{h}\right)$ such that $p_1\left(s^\rightarrow_{1}\right)\land\cdots\land p_n\left(s^\rightarrow_{n}\right)\rightarrow h\left(s^\rightarrow_{h}\right)$ must hold to make all non-fact rules in $\mathcal{H}$ SAT.
\end{definition}

A lemma can be directly derived from the definitions, signifying that a solution interpretation should summarize the information within the samples, and a sample should always evaluate to a certain truth value when making a valid hypothesis.
\begin{lemma}
\label{possamplemma}
If $\mathcal{H}$ is SAT, for each solution interpretation $\mathcal{I}^*$ of $\mathcal{H}$, 

(1) if $s^+$ is a positive sample of $p$ in $\mathcal{H}$,  we have $\mathcal{I}^*\left[p\right](s^+)=\top$.

(2) if $s^-$ is a negative sample of $p$ in $\mathcal{H}$,  we have $\mathcal{I}^*\left[p\right](s^-)=\bot$.

(3) if $s^{\rightarrow}=\left(s^\rightarrow_{1},\cdots,s^\rightarrow_{n},s^{\rightarrow}_{h}\right)$ is a implication sample of body predicates $(p_1,...,p_n)$ and head predicate $h$ in $\mathcal{H}$, $\mathcal{I}^*\left[p_1\right]\left(s^\rightarrow_{1}\right)\land\cdots\land \mathcal{I}^*\left[p_n\right]\left(s^\rightarrow_{n}\right)\rightarrow \mathcal{I}^*\left[h\right]\left(s^\rightarrow_{h}\right)$.
\end{lemma}

For ease of notation, for a sample $s$ and a CHC $\mathcal{C}$, we use $s[\mathcal{C}]$ to denote the clause $\mathcal{C}$ after a substitution, which replaces a list of variables with values when there is no ambiguity of the variable list.


\begin{example}
In $\mathcal{H}_0$ in Example \ref{exp}, for predicate $inv$, $s^+=(1,1,0,0,0)$ is a positive sample, since $s^+[\mathcal{C}_0]:\top\rightarrow inv(1,1,0,0,0)$ and $inv(1,1,0,0,0)=\top$ must hold to make $s^+[\mathcal{C}_0]$ SAT. $s^-=(2,1,5,0,5)$ is a negative sample, as $s^-[\mathcal{C}_2]:inv(2,1,5,0,5)\rightarrow \bot$. $s^\rightarrow=((1,1,0,0,0),(1,1,1,1,1))$ is an implication sample, because $s^\rightarrow[\mathcal{C}_1]:inv(1,1,0,0,0)\rightarrow inv(1,1,1,1,1)$.
\end{example}

We now formally define safe and unsafe zones, borrowing the terms ``safe'' and ``unsafe'' from program verification. 

\begin{definition}[Safe and Unsafe Zones]
\label{sudef}
A safe (unsafe) zone of a predicate $p$, $\mathcal{S}_p$ $(\mathcal{U}_p)$, is a set of positive (negative) samples of $p$. 
\end{definition}
The zones are often symbolically represented as expressions, such as inequalities and equations. They may include zero sample\footnote{In this case, the zone is $\bot$.}, or a finite or infinite number of samples. Notably, a positive sample can also be viewed as a safe zone, while a negative sample can be seen as an unsafe zone.

\begin{example}
\label{suzoneexp}
In $\mathcal{H}_0$ in Example \ref{exp}, one safe zone for predicate  $inv(v_0,v_1,v_2,v_3,v_4)$ is $\mathcal{S}_{inv}=\neg (v_0 \leq 0) \ \land \  v_1 \leq v_0 \ \land \ v_2=0 \ \land \ v_3=0 \ \land \ v_4=0$, since $s^+\in\mathcal{S}_{inv}$ satisfies $s^+[\mathcal{C}_0]:\top\rightarrow inv(s^+)$. One unsafe zone for predicate $inv$ is $\mathcal{U}_{inv}=\neg (v_4 \geq v_0 v_2)$, since $s^-\in\mathcal{U}_{inv}$ satisfies $s^-[\mathcal{C}_2]:inv(s^-)\rightarrow \bot $.
\end{example}


\subsection{Incorporate Zones within Learning Iterations}
\label{sec:chronosym}
\name is a framework that incorporates zones within the learning iterations.
A \textsl{Chronosymbolic Learner} proposes the hypothesis $\Tilde{\mathcal{I}}\left[p_i\right]$ considering currently reasoned safe zones $\mathcal{S}_{p_i}$, unsafe zones $\mathcal{U}_{p_i}$ from the reasoner with the inductive results (partitions) of the learner $\mathcal{L}_{p_i}$.

\begin{table}[]
\centering
\caption{Several candidates of making hypotheses.}
\label{tab:cand}
\renewcommand\arraystretch{1.4} 
\setlength{\tabcolsep}{3mm}{
\resizebox{\textwidth}{!}{
\begin{tabular}{ll}
\toprule
\textbf{Methods}  & \qquad\qquad\textbf{Candidate Hypothesis} 
\\ \midrule
BMC-styled & 

\begin{minipage}[c][0.7cm]{.55\textwidth}
\begin{flushleft}
\begin{equation}
\label{eq:s}
    \Tilde{\mathcal{I}}_{s}\left[p_i\right] =  \mathcal{S}_{p_i}
\end{equation}
\end{flushleft}
\end{minipage}
\\ \hline
LinearArbitrary-styled & 
\begin{minipage}[c][0.7cm]{.55\textwidth}
\begin{flushleft}
\begin{equation}
\label{l}
    \Tilde{\mathcal{I}}_{l}\left[p_i\right] =  \mathcal{L}_{p_i} \qquad
\end{equation}
\end{flushleft}
\end{minipage}
\\ \hline
Chronosymbolic w/o safe zones         &  
\begin{minipage}[c][0.7cm]{.55\textwidth}
\begin{flushleft}
\begin{equation}
\label{lu}
    \Tilde{\mathcal{I}}_{lu}\left[p_i\right] =  \mathcal{L}_{p_i} \land \neg\ \mathcal{U}_{p_i} \qquad
\end{equation}
\end{flushleft}
\end{minipage}
\\ \hline
Chronosymbolic w/o unsafe zones         &         
\begin{minipage}[c][0.7cm]{.55\textwidth}
\begin{flushleft}
\begin{equation}
\label{sl}
    \Tilde{\mathcal{I}}_{sl}\left[p_i\right] = \mathcal{S}_{p_i} \lor  \mathcal{L}_{p_i} \qquad
\end{equation}
\end{flushleft}
\end{minipage}
\\ \hline
Chronosymbolic          &          
\begin{minipage}[c][0.7cm]{.55\textwidth}
\begin{flushleft}
\begin{equation}
\label{slu}
    \Tilde{\mathcal{I}}_{slu}\left[p_i\right] = \mathcal{S}_{p_i} \lor \left( \mathcal{L}_{p_i} \land \neg\ \mathcal{U}_{p_i} \right) \quad
\end{equation}
\end{flushleft}
\end{minipage}
\\ \bottomrule
\end{tabular}
}
}
\vspace{-0.5cm}
\end{table}

Several promising candidates of the hypothesis $\Tilde{\mathcal{I}}\left[p_i\right]$ that can be made by \textsl{Chronosymbolic Learner}\footnote{This is the procedure \texttt{makeHypothesis()} in Algorithm \ref{alg}.} are listed in Table \ref{tab:cand}.
Note that Equation (\ref{slu}) simply serves as an example of a Chronosymbolic hypothesis. It can be replaced by any symbolic classification algorithm taking samples and zones as input, providing a symbolic hypothesis as output.
We have the following lemma: 


\begin{lemma}
\label{chrohypo}
$\Tilde{\mathcal{I}}_{slu} \succeq \{\Tilde{\mathcal{I}}_{sl},\Tilde{\mathcal{I}}_{lu}\} \succeq \Tilde{\mathcal{I}}_{l}$, where
 $A \succeq B$ denotes that A is a solution interpretation of the CHC system $\mathcal{H}$ whenever B is a solution interpretation. 
\end{lemma}

Lemma \ref{chrohypo} shows the order of the feasibility of becoming a solution interpretation.
We give a proof sketch for $\Tilde{\mathcal{I}}_{sl} \succeq \Tilde{\mathcal{I}}_{l}$ and proofs for others are similar.
\label{proofsket}
\vspace{-0.5cm}
\begin{proof}
Assume $\Tilde{\mathcal{I}}_{l}\left[p_i\right]=\mathcal{L}_{p_i}$ is a solution interpretation of $\mathcal{H}$, and then by Definition \ref{possamp}, we have for each positive sample $s^+$, $\Tilde{\mathcal{I}}_{l}\left[p_i\right](s^+)=\top$, i.e., $s^+ \in \Tilde{\mathcal{I}}_{l}\left[p_i\right]$. 
According to Definition \ref{sudef}, the safe zone $\mathcal{S}_{p_i}$ is a set of positive samples, and we have $\mathcal{S}_{p_i} \subseteq \Tilde{\mathcal{I}}_{l}\left[p_i\right]$. We now have $\Tilde{\mathcal{I}}_{sl}\left[p_i\right]=\mathcal{S}_{p_i} \lor \Tilde{\mathcal{I}}_{l}\left[p_i\right]=\Tilde{\mathcal{I}}_{l}\left[p_i\right]$ and $\Tilde{\mathcal{I}}_{sl}\left[p_i\right]$ is also a solution interpretation. 
Thus, we can conclude $\Tilde{\mathcal{I}}_{sl} \succeq \Tilde{\mathcal{I}}_{l}$.\qed
\end{proof}
In practice, Equation (\ref{slu}) generally delivers the optimal result among the candidates, as it encapsulates more information summarized by both the learner and the reasoner\footnote{Appendix \ref{app:sltana} provides additional theoretical analysis on why \name performs better from the perspective of the state and solution space.}. 
Nevertheless, this does not apply to every individual instance, since the introduction of zones might alter the exploration of the state space\footnote{This depends on the algorithms used in the teacher (in our instance, an SMT solver) to get counterexamples. After incorporating zones, the teacher may also return counterexamples that lead to less progress for the learner.}. To address this issue, we can manually apply adequate strategies on each instance, or use a \textit{scheduler} to alternate those candidate hypotheses over time. Our experiments will demonstrate the performance improvement achieved by employing various strategies.

\subsection{Overall Algorithm}
\label{algor}
\renewcommand{\algorithmicrequire}{\textbf{input:}}  
\renewcommand{\algorithmicensure}{\textbf{output:}} 

{
\linespread{1.1}
\begin{algorithm}[ht!]
    \caption{\name $\left(\mathcal{H}\right)$} 
        \label{alg}
    \begin{algorithmic} [1]
        \REQUIRE A CHC system $\mathcal{H}=\{\mathcal{C}_0,\cdots,\mathcal{C}_n\}$
        \STATE $\bullet$ Initialize safe and unsafe zones \qquad $\forall \  p_i, \ (\mathcal{S}_{p_i}, \mathcal{U}_{p_i})$ $\leftarrow$ $(\bot,\bot)$
        \STATE $\bullet$ Initialize the hypotheses \qquad \qquad \quad $\forall \  p_i, \ \Tilde{\mathcal{I}}_{p_i}$  $\leftarrow$ $\top$ 
        \STATE  $\bullet$ Initialize UNSAT flag \qquad \qquad \qquad  $is\_unsat$ $\leftarrow \bot$ 
        \STATE  $\bullet$ Initialize the dataset \qquad \qquad \qquad \ \   $\mathcal{D}\leftarrow \varnothing$ 
            \WHILE{not $is\_unsat$ and $\exists \ \mathcal{C}_i,$ not \texttt{SMTCheck(}$\Tilde{\mathcal{I}}\left[\mathcal{C}_i\right]$\texttt{)}}
                \STATE $\circ$ Reasoning and UNSAT checking\  $\ (\mathcal{S}$, $\mathcal{U}$, $is\_unsat, \Tilde{\mathcal{I}},\mathcal{R})$ $\leftarrow$\ \texttt{reason(}$\mathcal{H}$, $\mathcal{S}$, $\mathcal{U}$\texttt{)}
                \STATE $\circ$ Sample from zones and add to $\mathcal{D}$ \quad $\mathcal{D}\leftarrow\mathcal{D}\ +$ \texttt{sampling(}$\mathcal{S}$, $\mathcal{U}, \mathcal{D}$) 
                
                \FOR{each $\mathcal{C}_i \in \mathcal{H}$}
                \WHILE{not \texttt{SMTCheck(}$\Tilde{\mathcal{I}}\left[\mathcal{C}_i\right]$\texttt{)}}
                
                \STATE $\bullet$ Find counterexample(s) \qquad \ $c\leftarrow$  \texttt{SMTModel(}$\Tilde{\mathcal{I}}\left[\mathcal{C}_i\right]$\texttt{)}
                \STATE $\bullet$ Convert counterexample(s) to samples \qquad $s\leftarrow$ \texttt{converting(}c\texttt{)}
                \STATE $\bullet$ UNSAT checking \qquad \qquad \quad $is\_unsat, \Tilde{\mathcal{I}}, \mathcal{R}$ $\leftarrow$ \texttt{checkUNSAT($\mathcal{D}$, s)}
                \STATE $\bullet$ Add samples to $\mathcal{D}$ \qquad \qquad \ $\mathcal{D}\leftarrow \mathcal{D}+s$, if not $is\_unsat$

                \STATE $\bullet$ Learn a partition \qquad \qquad \quad $\mathcal{L}\leftarrow$ \texttt{learn(}$\mathcal{D}$\texttt{)}
                \STATE $\bullet$ Update the hypothesis \qquad \ \ $\Tilde{\mathcal{I}}\leftarrow$ \texttt{makeHypothesis(}$\mathcal{S}, \mathcal{U}, \mathcal{L}$\texttt{)}
                \ENDWHILE
                
                \ENDFOR
               
            \ENDWHILE
        \ENSURE A solution ($is\_unsat,\ \Tilde{\mathcal{I}}, \mathcal{R}$) for the CHC system $\mathcal{H}$
    \end{algorithmic}
\end{algorithm}  

}

The overall algorithm is outlined in Algorithm \ref{alg}, where the solid bullet points denote mandatory steps and the unfilled bullet points indicate optional steps, suggesting that they need not be executed in every iteration.

The algorithm takes a CHC system $\mathcal{H}$ as input and outputs a solution for it. Initialization of zones, hypotheses, UNSAT flag, and the dataset occurs in lines 1-4.
The outer while loop, spanning lines 5 to 18, checks if a solution of $\mathcal{H}$ has been found. If not, it initiates a new epoch and continues solving. In line 5, \texttt{SMTCheck(}$\Tilde{\mathcal{I}}\left[\mathcal{C}_i\right]$\texttt{)} calls the backend SMT solver to check the satisfiability of a CHC $\mathcal{C}_i$ under current hypothesis interpretation $\Tilde{\mathcal{I}}$. 
In line 6, the reasoner refines the zones and checks if the zones overlap (see Section \ref{sec:reasoner}). 
Line 7 involves the \textit{sampling} procedure described in Section \ref{sec:sample}.

The \textbf{for} loop from lines 8-17 iteratively finds an UNSAT CHC and refines the hypothesis until it becomes SAT\footnote{An alternative design choice is to update the hypothesis only once and move to the next UNSAT CHC, which achieves better results in some cases.}.
In the inner while loop in lines 10-13, we find one or a batch of counterexamples, convert them into samples, and add them to the dataset after the UNSAT checking\footnote{For simplicity, the refutation proof generated when checking UNSAT is also represented as $\Tilde{\mathcal{I}}$ in Algorithm \ref{alg}.} passes. 
\texttt{SMTModel(}$\Tilde{\mathcal{I}}\left[\mathcal{C}_i\right]$\texttt{)} in line 10 calls the backend SMT solver to return a counterexample that elucidates why the current hypothesis is invalid.
Following dataset updates, in lines 14-15, the learner induces a new partition. We then update the Chronosymbolic hypothesis by combining the zonal information and the partition, as described in Section \ref{sec:chronosym}. 
The new hypothesis is more likely to be a solution interpretation (if the CHC system is SAT), as it integrates the newly acquired information gathered in this iteration\footnote{Here we do not consider approximation error in classification or tentative samples defined later.}.

\section{Design of Learner and Reasoner}
In this section, we provide a simple instance of \name, utilizing standard procedures as illustrative examples to demonstrate how they can be integrated into our framework. Implementation details can be found in Appendix \ref{sec:impldet}.
Discussion on alternative design choices is provided in Section \ref{sec:discussion}.
\label{sec:design_of_lr}
\subsection{Learner: Data-Driven CHC Solving}
\label{sec:learner}

The \textit{learner} module in our framework leverages an induction-based and CEGAR-inspired \cite{cegar} CHC solving scheme.
It has two sub-modules: the \textit{dataset} and the \textit{machine learning toolchain}. The dataset stores the positive and negative samples and the corresponding predicates for inductive learning. The machine learning toolchain (an example introduced in Section \ref{mltool}) takes the samples in the dataset as input and outputs a \textit{partition} that correctly classifies all these samples.

\subsubsection{Converting: From Counterexamples to Samples.}
\label{sec:converting}
The positive and negative samples in our framework are \textit{converted} from \textit{counterexamples}. As in Fig. \ref{arc}, the counterexample provided by the teacher is essential in the learning loop, since it offers the learner information about why the current hypothesis is incorrect and how to improve it. 
Here we formally define the term ``counterexample''.

\begin{definition}[Counterexample]
\label{cex}
A counterexample $c=\bigl((s_{p_1},\cdots,s_{p_k}),s_{h}\bigr)$ for a CHC $\mathcal{C}$ and an interpretation $\mathcal{I}$
is a set of data points\footnote{They can also be extended to zonal representation, with a possible approach described by \cite{xu2020interval}.}
such that under $c$, $\mathcal{I}\left[\mathcal{C}^\prime\right]=\bot$, where $\mathcal{C}^\prime$ is the same constraint as $\mathcal{C}$ but without the quantifier.
\end{definition}
\begin{example}
In $\mathcal{H}_0$ in Example \ref{exp}, given that the current hypothesis of predicate $inv(v_0,v_1,v_2,v_3,v_4)$ is $\Tilde{\mathcal{I}}\left[inv\right]=v_3\leq1$, a counterexample for $\mathcal{C}_1$ could be $c=(((2,0,0,0,0)),(2,0,1,2,0))$, as it makes $\Tilde{\mathcal{I}}[\mathcal{C}_1]:\top\rightarrow\bot$.
\end{example}

Counterexamples can be converted into positive, negative, or implication samples through the \textit{converting} procedure in Fig. \ref{arc} and Algorithm \ref{alg}.
The following converting scheme maps counterexamples into positive, negative, and implication samples based on the CHC type. Essentially, this approach samples the CHC system in three directions: forward, backward, and middle-out.

\begin{lemma}[Sample Converting]
\label{factsample}
A fact's counterexample can provide a positive sample for the head predicate.
A linear query's counterexample can provide a negative sample for the body predicate.
A non-fact rule's counterexample can provide an implication sample of the body predicate(s) and the head predicate.
\end{lemma}

To simplify the problem into a classic binary classification format and enable off-the-shelf machine learning tools to handle implication samples more effectively, we introduce ``tentative sample\footnote{It is extensible to \textit{tentative zones} to make an IC3-styled reasoner possible.}'':

\begin{definition}[Tentative Sample]
A tentative positive (negative) sample $\Tilde{s}^+$ $(\Tilde{s}^-)$ of predicate $p$ is a data point such that, 
the learner tentatively deems it to be a positive (negative) sample, but 
it may not satisfy Definition \ref{possamp} (\ref{negsamp}).
\end{definition}

\begin{lemma}
[Implication Sample Converting]
For an implication sample $s^\rightarrow$, if the $n$ data points in $(s^\rightarrow_{1},\cdots, s^\rightarrow_{n})$ are all positive samples (or in a safe zone), then $s^{\rightarrow}_{h}$ is a positive sample. If the data point $s^{\rightarrow}_{h}$ is a negative sample (or in an unsafe zone), and $n=1$, the data point $s^\rightarrow_{1}$ is a negative sample. Otherwise, the sample can only be converted into tentative positive or negative samples, whose tentative categories make the logical implication hold.
\end{lemma}

Empirically, converting all tentative samples into negative tentative samples is an appropriate design choice.
As these samples are not guaranteed positive or negative, periodic clearing is necessary to avoid keeping the ``wrong guess'' in a specific tentative category forever. 
Following this conversion, only positive and negative samples remain visible to the learner, enabling natural adaptation with off-the-shelf machine learning tools for binary classification like DT and SVM.

\subsubsection{Sampling: Obtain Data Points from Zones.}
\label{sec:sample}
To find extra data points and maximize the usage of the zones, we provide an alternative data collection strategy, i.e., \textit{sampling} data points from zones.
This procedure involves selecting a sample from a given zone.
Typically, it calls the backend SMT solver with a safe (unsafe) zone as the input constraint. The SMT solver returns positive (negative) data points within the zone while ensuring the avoidance of duplicates already present in the dataset.

\subsubsection{UNSAT Checking.}
Two lemmas show when the learner determines a CHC system as UNSAT. Note that tentative samples are not used in UNSAT checking.
\begin{lemma}[Sample-Sample Conflict]
\label{lemma:ssc}
If there exists a sample $s$ for $p$ in $\mathcal{H}$ that is both a positive and a negative sample simultaneously, then $\mathcal{H}$ is UNSAT.
\end{lemma}
\begin{proof}
According to Lemma \ref{possamplemma}, if $s$ is both a positive and negative sample, suppose $\mathcal{H}$ is SAT with a solution interpretation $\mathcal{I}^*$, we have $\mathcal{I}^*\left[p\right](s)=\top$ and $\mathcal{I}^*\left[p\right](s)=\bot$, which contradict each other. Hence, $\mathcal{H}$ is UNSAT. \qed
\end{proof}

\begin{lemma}[Sample-Zone Conflict]
\label{lemma:szc}
If there exists a sample $s$ for $p$ in $\mathcal{H}$ that is a positive sample and is in an unsafe zone, or is a negative sample and is in a safe zone, then $\mathcal{H}$  is UNSAT.
\end{lemma}
\begin{proof}
With Definition \ref{sudef}, sample-zone conflict can be reduced to sample-sample conflict, the case in Lemma \ref{lemma:ssc}.\qed
\end{proof}

\subsection{Reasoner: Zones Discovery}
\label{sec:reasoner}
For the \textit{reasoner}, we adopt a simple BMC-styled
logical reasoning procedure for inferring safe and unsafe zones in our instance of \name. 
The role of the reasoner is not to deduce a complete hypothesis but to furnish a ``partial solution'' in the form of safe and unsafe zones,  collaborating well with the learner. 
The zones condense information within a bounded number of transitions.
In our tool, the reasoner initializes the zones, expands the zones through image (or pre-image) computation, and does UNSAT checking. 

In each of the subsequent lemmas, all variables utilized in the discussed CHC are denoted as $\mathcal{X}$. Among them, variables used by the discussed predicate are denoted as $\mathcal{X}_{\mu} \subseteq \mathcal{X}$. The other variables are called \textit{free variables}, denoted as $\mathcal{X}_{\varphi} = \mathcal{X} \setminus \mathcal{X}_{\mu} \subseteq \mathcal{X} $.

\subsubsection{Initialization of Zones.}


We provide lemmas about initiating zones. 

\begin{lemma}[Initial Safe Zones]
A fact $\mathcal{C}_f: \phi_f\rightarrow h_f(T)$ produces a safe zone for $h_f$: $\mathcal{S}^0_{h_f}(T)=\exists \mathcal{X}_{\varphi}, \phi_f$.
\end{lemma}

\begin{lemma}[Initial Unsafe Zones]
A linear query $\mathcal{C}_q: \phi_q \land p_q(T)\rightarrow \bot$  produces an unsafe zone for $p_q$: $\mathcal{U}^0_{p_q}(T)=\exists \mathcal{X}_{\varphi}, \phi_q$.
\end{lemma}

\subsubsection{Expansion of the Zones.}
\label{sec:expansion}
Zones can be expanded in the following ways:
\begin{lemma}[Forward Expansion]
From given safe zones $\mathcal{S}^m_{p_i}$,  we can expand them in one forward transition by a non-fact rule $\mathcal{C}_r:  \phi  \land p_1(T_1) \land \cdots \land p_k(T_k) \rightarrow h(T)$ to get an expanded safe zone $\mathcal{S}^{m+1}_{h}$,  where $\mathcal{S}^{m+1}_{h}(T)=\exists \mathcal{X}_{\varphi},\phi\land\mathcal{S}^m_{p_1}(T_1)\land\cdots \land {S}^m_{p_k}(T_k)$.
\end{lemma}

\begin{lemma}[Backward Expansion]
From a given unsafe zone $\mathcal{U}^m_{h}$, we can expand it in one backward transition by a non-fact linear rule $\mathcal{C}_r:  \phi  \land p(T_0) \rightarrow h(T)$ to get an expanded unsafe zone $\mathcal{U}^{m+1}_{p}$,  where $\mathcal{U}^{m+1}_{p}(T_0)=\exists \mathcal{X}_{\varphi},\phi\land\mathcal{U}^{m}_{h}$.
\end{lemma}

In the context of program verification, the forward expansion incorporates more information about the reached states.
The backward expansion expands the set of states under which the program is deemed unsafe when reached. Note that the expansion operation can be computationally expensive, and excessive expansion leads to a heavy burden for the backend SMT solver. 

\subsubsection{UNSAT Checking.}
We further discuss when the reasoner determines UNSAT.
\begin{lemma}[Zone-Zone Conflict]
\label{lemma:zzc}
If there exists a predicate $p$ in $\mathcal{H}$  that its safe zone $\mathcal{S}_p$ and unsafe zone $\mathcal{U}_p$ overlap, i.e., $\mathcal{S}_p \land \mathcal{U}_p$ is SAT, then $\mathcal{H}$ is UNSAT.
\end{lemma}
\begin{proof}
According to Definition \ref{sudef}, we sample one data point from the overlap of zones, and we can then apply Lemma \ref{lemma:szc} to prove Lemma \ref{lemma:zzc}.\qed
\end{proof}
\section{Experiments}
\label{sec:experiment}
\subsection{Experimental Settings}
\label{expsetting}
\paragraph{Benchmarks.} Our benchmarking dataset comprises a collection of widely-used arithmetic CHC benchmarks from related works \cite{dillig2013inductive,freq,garg2016learning} and SV-COMP\footnote{\url{http://sv-comp.sosy-lab.org/}}, collected by FreqHorn \cite{freq}. It is a suite of $288$ instances in total, consisting of $235$ safe ones and $53$ unsafe ones. The instances are lightweight arithmetic CHC systems, yet many demand non-trivial effort to determine a solution interpretation due to the arithmetic complexity. 
Most instances are under linear arithmetic, while a few necessitate handling non-linear arithmetic. 
We also discuss results on CHC-COMP-22\footnote{https://chc-comp.github.io/}, LIA track.
See Appendix \ref{app:statbench} for details.

\paragraph{Hyperparameters.} 
In SVM, the $C$ parameter is set to $1$. The upper limit of the coefficients in the hyperplanes is set to $5$. 
For DT, we use C5.0 as a default option.
For the dataset, the best result we achieved is to use option B in Appendix \ref{sec:learner_ext}, where $a=50$, $b=50$.
For the reasoner, the hyperparameters $l_1$, $l_2$ and $l_3$ in Appendix \ref{ab} are $700$, $500$, and $1500$\footnote{More information on the configuration, like the scheduler mentioned in Section~\ref{sec:chronosym}, is shown at \url{https://github.com/Chronosymbolic/Chronosymbolic-Learning/blob/main/experiment/result_safe_summary.log}.} respectively.

\paragraph{Experiment Setup.} We impose a timeout of solving a CHC system instance as $360$ seconds across all tools. To account for the inherent randomness in the tools, we repeat three times for each experiment and report the best result\footnote{We include results showcasing how random seeds affect the performance in Appendix \ref{app:rnd}.}.

\subsection{Baselines}
State-of-the-art CHC solvers can be classified into three categories: reasoning-based, synthesis-based, and induction-based solvers (see Section \ref{relwork}). We compare our method against the representatives of each type as follows:

\paragraph{LinearArbitrary.} LinearArbitrary \cite{zhu2018data} is the state-of-the-art induction-based solver that learns inductive invariants from counterexample-derived samples using SVM and DT. We use the default hyperparameters in the paper and code\footnote{\url{https://github.com/GaloisInc/LinearArbitrary-SeaHorn}}. 

\paragraph{FreqHorn.} FreqHorn\footnote{\url{https://github.com/freqhorn/freqhorn}} \cite{freq} is a representative of synthesis-based solvers, synthesizing invariants using frequency distribution of patterns. It cannot determine the CHC system as UNSAT, and it resorts to a separate procedure ``Freqhorn-expl'' to do that. In the experiment, we adhere to the default settings in \cite{freq}.

\paragraph{Spacer and GSpacer.} 
Spacer \cite{komuravelli2013automatic} is a powerful SMT-based CHC-solving engine serving as the default CHC engine in Microsoft Z3\footnote{\url{https://github.com/Z3Prover/z3}} \cite{de2008z3}. 
The experiment is conducted under the default settings of Microsoft Z3 4.8.11.0.
GSpacer\footnote{\url{https://github.com/hgvk94/z3/commits/ggbranch}} \cite{vediramana2020global} achieves state-of-the-art on many existing CHC benchmarks by extending with some well-designed global guidance heuristics. To achieve the best result, we enable all the heuristics, including the subsume, concretize, and conjecture rules.

\subsection{Performance Evaluation}

\begin{wrapfigure}{r}{5.5cm}
\vspace{-1.5cm}
\centering
\includegraphics[width=0.5\textwidth]{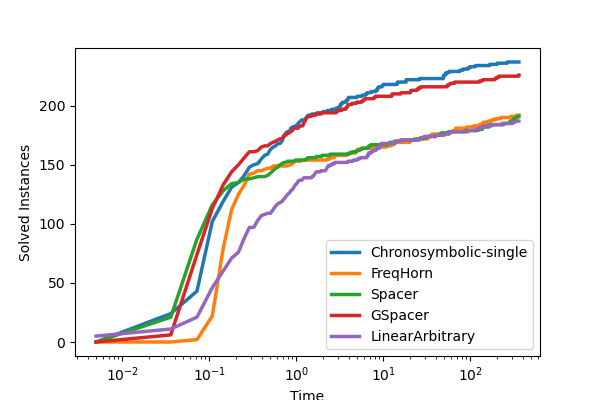}
\caption{
The solved instances for different methods as the running time increases.
}
\label{fig:eff}
\end{wrapfigure}

\begin{figure}[h]
\centering
\includegraphics[width=\textwidth]{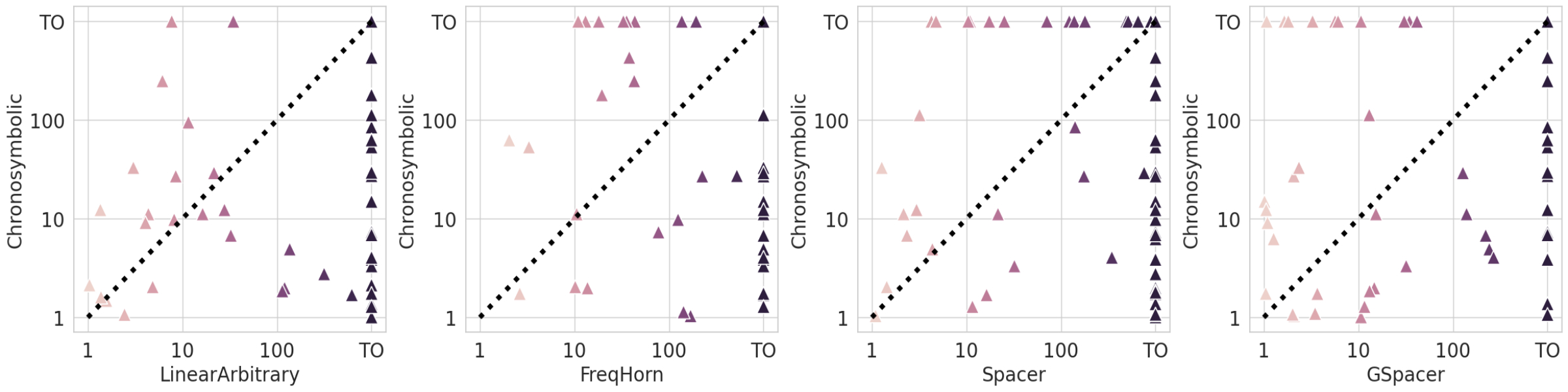}
\caption{
Running time comparison of \name and baselines, where the timeout (TO) is 360 seconds. The points below the diagonal indicate instances on which Chronosymbolic-single outperforms baselines. Here, we only plot non-trivial instances (i.e., solving time takes $ \geq 1 $ second for both solvers). 
}
\label{fig:comparison}
\end{figure}

The performance comparison \footnote{For detailed timing of each instance, see \url{https://github.com/Chronosymbolic/Chronosymbolic-Learning/blob/main/experiment/comparison.xlsx}} is shown in Table \ref{tab:res}.
``Chronosymbolic-single'' configuration stands for the best result from a single run of our solver, where all instances are tested using a fixed set of hyperparameters. 
``Chronosymbolic-cover'' encompasses all solved benchmarks achieved through 13 runs\footnote{The details are included in Appendix \ref{app:detcover}.} using different strategies and hyperparameters, demonstrating the potential capability of our framework when selecting an adequate configuration for each instance. Timing information for this setting is not included because the time to solve a certain benchmark varies among different runs.
Fig. \ref{fig:eff} shows the solved instances for compared methods as the running time increases, and Fig. \ref{fig:comparison} shows the runtime comparison per instance.

\begin{table}[]
\centering
\vspace{-0.6cm}
\renewcommand\arraystretch{1.2}
\caption{Performance evaluation. ``\#total'' stands for the total number of solved instances and ``percentage'' for the percentage of solved ones among all 288 instances, ``\#safe'' and ``\#unsafe'' for solved ones among 235 safe instances and 53 unsafe instances respectively. For timing, ``avg-time'' is the average time consumed on each instance (including timeout or crashed ones), and ``avg-time-solved'' is the average time consumed on each solved instance.}
\label{tab:res}
\vspace{0.1cm}
\resizebox{\textwidth}{!}{
\setlength{\tabcolsep}{1.5mm}{
\begin{tabular}{cccccccc}
\hline
            Method    & \#total & percentage & \#safe & \#unsafe & avg-time (s) & avg-time-solved (s) \\ \hline
LinearArbitrary & 187     & 64.93\%    & 148    & 39       &  135.0   &    13.48      \\
FreqHorn        & 191     & 66.32\%    & 191    & 0        &    129.1  &    11.80       \\
FreqHorn-expl        & 50     & 17.36\%    & 0    & 50        &    299.5  &    13.57       \\
Spacer          & 184     & 63.89\%    & 132    & 52       & 132.8  &    15.30   \\
GSpacer         & 220     & 76.39\%    & 174    & 46       & 83.50  &    7.83   \\
\hline
\textbf{Chronosymbolic-single} & \textbf{237} & \textbf{82.29\%} & \textbf{189} & \textbf{48} & \textbf{68.33}  & \textbf{7.51}\\
\textbf{Chronosymbolic-cover}  & \textbf{252} & \textbf{87.50\%} & \textbf{204} & \textbf{48} & \textbf{-}   &  \textbf{-} \\ \hline
\end{tabular}}
}
\vspace{-0.5cm}
\end{table}

Our tool, even in the Chronosymbolic-single setting, solves $17$ more instances than the best competing solver, GSpacer, and generally outperforms other solvers in terms of speed, even though it is implemented in Python. 
This result shows that our tool performs considerably better even without careful tuning for each CHC instance. 
In the Chronosymbolic-cover setting, our approach shows significant improvements, emphasizing the need for tailored strategies for different instances.
Among other approaches, GSpacer performs the best on solved instances and time consumed, as reported in prior literature. Nevertheless, Spacer, GSpacer, and LinearArbitrary struggle with most instances involving non-linear arithmetic. FreqHorn can solve some of them, but fails in all unsafe instances without resorting to another procedure ``expl''. Our approach can also handle limited non-linearity because our reasoner can induce non-linear zones. 

From Fig. \ref{fig:eff}, we note that Chronosymbolic-single outperforms FreqHorn and LinearArbitrary under any timeout within 360 seconds, showing the enhanced efficiency of our method. It is also expected that Spacer and GSpacer perform the best under an extremely short time limit as they do not involve time-consuming inductive learning.

\begin{wraptable}{r}{5.7cm}
\centering
\vspace{-1.2cm}
\renewcommand\arraystretch{1.1}
\caption{Evaluation on CHC-COMP-2022-LIA. ``\#total'' and ``percentage'' stand for the total number and the percentage of solved instances among 499 instances respectively. }
\label{tab:res_chccomp}
\vspace{-0.2cm}
\resizebox{0.45\textwidth}{!}{
\setlength{\tabcolsep}{1.5mm}{
\begin{tabular}{ccc}
\hline
            Method    & \#total & percentage  \\ \hline
LinearArbitrary & 156     & 31.26\%         \\
FreqHorn        & 123     & 24.65\%           \\
Spacer          & 261     & 52.30\%      \\
GSpacer         & 318     & 63.73\%      \\
\hline
\begin{tabular}{@{}c@{}}
Chronosymbolic-single
\end{tabular}
 & 197 & 39.48\% \\
\hline
\end{tabular}}
}
\vspace{-1cm}
\end{wraptable}
For CHC-COMP, most benchmarks represent large transition systems with a substantial number of Boolean variables. Reasoning-based approaches inherently suit these benchmarks better, and unsurprisingly, invoking the machine learning toolchain in each iteration is an inefficient design choice. 
An experiment shows that by filtering out benchmarks in CHC-COMP-22-LIA with excessively large sizes (e.g., exceeding 100 rules or 200 variables), our results are on par with the state-of-the-art (Chronosymbolic-cover 129/208 vs. GSpacer 130/208).

\subsection{Ablation Study}
We further study the effectiveness of the critical parts of our framework: the learner and the reasoner.
For the reasoner, we compare the result 
when safe zones are not provided, when unsafe zones are not provided, and when neither are provided. 
Additionally, we assess the performance when entirely removing the learner and running our tool solely with the reasoner.
Throughout this ablation study, all other settings remain consistent with  Chronosymbolic-single.

\begin{table}[]
\vspace{-0.7cm}
\centering
\caption{Different configurations of \name.}
\renewcommand\arraystretch{1.2}
\resizebox{\textwidth}{!}{
\setlength{\tabcolsep}{1.5mm}{
\begin{tabular}{cccccccc}
\hline
Configuration                  & \#total      & percentage       & \#safe       & \#unsafe    & avg-time (s)   & avg-time-solved (s) \\ \hline
without safe zones   & 228 & 79.17\% & 183 & 45 & 78.56 & 9.11 \\
without unsafe zones & 218 & 75.69\% & 173 & 45 &  94.75 & 12.87 \\
without both zones   & 211 & 73.26\% & 166 & 45 &  93.84  & 10.09 \\
without learner      & 131    &  45.49\%       &  96   &  35  & 196.3  &  0.16 \\
parallel      & 216    &  75.00\%       &  180   &  36  & -  &  - \\\hline
\textbf{Chronosymbolic-single} & \textbf{237} & \textbf{82.29\%} & \textbf{189} & \textbf{48} & \textbf{68.33} & \textbf{7.51}       \\ \hline
\end{tabular}}}
\label{tab:ablation}
\vspace{-0.5cm}
\end{table}

From Table \ref{tab:ablation}, as expected, Chronosymbolic-single comprehensively outperforms all other configurations, which shows that each component of our framework is indispensable. It is also clear that each zone provides useful information to the hypothesis, making the result of Chronosymbolic without both zones worse than without safe and unsafe zones. 
As discussed in Section \ref{sec:expansion} and Appendix \ref{sec:reasoner_ext}, our lightweight reasoner is intended to aid the learner. Thus, not utilizing a learner yields the poorest result, but at a faster computation speed.
The result also shows that our learner\footnote{``Without both zones'' indicates that the reasoner does not participate.} performs much better than LinearArbitrary, as we have a more comprehensive converting and sampling scheme.

Another experiment in Table \ref{tab:ablation}, ``parallel'', reveals that if we run our learner and reasoner \textit{individually and simultaneously} for $360$ seconds, they only solve a total of $180$ safe and $36$ unsafe instances. This number is significantly lower than what our proposed method achieves, which underscores the mutual benefit of the reasoner and learner in \name.

\section{Related Work}
\label{relwork}
CHC solving has garnered extensive research, offering numerous artifacts for applications like software model checking, verification, and safe inductive invariant generation. Modern CHC solvers primarily rely on three categories of techniques\footnote{See Appendix \ref{extrelwork} for details.}. 
1) \textit{Symbolic Reasoning.}
Solvers in this category \cite{bmc, mcmillan2003interpolation, hojjat2018eldarica, bradley2011sat, vizel2014interpolating, komuravelli2013automatic, vediramana2020global} maximize the power of logical reasoning and utilize a series of heuristics to accelerate the reasoning engine.
2) \textit{Synthesis. }
Methods in this category \cite{si2020code2inv, freq, dillig2013inductive, yao2020learning, riley2022multi, padhi2016data} typically reduce the CHC solving problem into an invariant synthesis problem, which aims to construct an inductive invariant under the \textit{semantic} and \textit{syntactic} constraints.
3) \textit{Induction.}
Instead of explicitly constructing the interpretation of unknown predicates, induction-based CHC solvers \cite{garg2014ice,champion2018hoice, garg2016learning, xu2020interval, zhu2018data, salzberg1994c4, cegar, nguyen2017counterexample, sharma2013data} directly learn such an interpretation by induction learning from data.

\section{Discussion and Future Work}
\label{sec:discussion}
\paragraph{Discussion on the alternative design choices for learner and reasoner.} Many existing methods can potentially be integrated into our modular framework.
For the learner in \name, ICE framework \cite{garg2014ice}, ICE-DT \cite{garg2016learning} Interval-DT \cite{xu2020interval} which have special designs for implication samples can also be valid choices.
Additionally, symbolic (binary) classification algorithms can be a suitable choice.
For the reasoner, model-based projection (MBP) \cite{komuravelli2016smt} can be integrated straightforwardly, while IC3 \cite{bradley2011sat}, PDR \cite{vizel2014interpolating}, Spacer \cite{komuravelli2013automatic}, and GSpacer \cite{vediramana2020global} require more careful design (e.g., supporting tentative zones) because of the introduction of over-approximation zones.
If such tentative zones are introduced, a wider range of methods that generalize data points to zones, e.g., symbolic regression \cite{makke2024interpretable,brolos2021approach} methods and convex hull algorithms \cite{bjorner2015property} can be applied.

\paragraph{Future work.}
Despite the promising result, there are several avenues for future work. 
Firstly, our learner primarily generates linear expressions (and some simple non-linear operators like \texttt{mod}), necessitating adaptation for broader abstract domains. 
It is also crucial to develop and integrate machine learning algorithms that are specifically tailored for this problem within the framework (e.g., how we can obtain a good partition or classifier with the existence of samples and zones).
Second, our strategy for identifying safe and unsafe zones is primitive, prompting the exploration of advanced logical reasoning algorithms like IC3-style approaches, where zone generalization and summarization should be carefully balanced. 
Thirdly, we acknowledge the influence of different learning strategies on benchmark solving (as Chronosymbolic-cover outperforms Chronosymbolic-single by a large margin) and intend to conduct comprehensive analyses for automated strategy selection.
Lastly, currently, our tool only uses elementary algorithms for zone reasoning, which is not optimized for large CHC systems (e.g., many instances in CHC-COMP). Additionally, our algorithmic support for non-linear CHCs is also limited.
We plan to dedicate efforts to engineering improvements and explore ways to enhance efficiency in downstream machine learning procedures, particularly when handling numerous Boolean variables.

\section{Conclusion}
In this work, we propose \name, a framework that can synergize reasoning-based techniques with data-driven CHC solving in a reciprocal manner.
We also provide a simple yet effective instance of \name to demonstrate its functionality and showcase its potential.
Our experiments, conducted on 288 commonly used arithmetic CHC benchmarks, reveal that our tool outperforms several state-of-the-art CHC solvers, encompassing reasoning-based, synthesis-based, and induction-based methods.


\section*{Acknowledgment}
We sincerely thank the reviewers for their insightful questions and comments. 
We would like to thank Hari Govind Vediramana Krishnan, and Arie Gurfinkel for discussing and providing information about Z3 solver, Spacer and GSpacer, Adam Weiss and Zhixin Xiong for conducting initial work, Breandan Considine for proofreading, Zhaoyu Li and Chuqin Geng for sharing ideas, and Doina Precup for supervision. 


\bibliographystyle{splncs04}
\bibliography{main}

\newpage
\appendix
\clearpage
\section*{Appendix}

\section{Extended Preliminary}
\label{app:extpre}
\subsection{Extended Related Work}
\label{extrelwork}
Modern CHC solvers primarily rely on the following three categories of techniques.
\paragraph{Symbolic Reasoning.}
Solvers in this category maximize the power of logical reasoning and utilize a series of heuristics to accelerate the reasoning engine. The classic methods are mostly SAT-based.
\textit{Bounded Model Checking} (BMC) \cite{bmc} encodes the initial states, transitions, and bad states into logical variables, and unfolds $k$ transitions to form a logical formula. 
However, it only provides a \textit{bounded} proof that can only show the correctness within a certain number of transitions.
\textit{Craig Interpolation} (CI) \cite{mcmillan2003interpolation} tries to find an ``interpolant'' that serves as an over-approximating summary of reachable states. If such an interpolant is a fixed-point, then an \textit{unbounded} proof (inductive invariant) is found. \textit{ELDARICA} \cite{hojjat2018eldarica} is an instance of maximizing the power of CI, as it constructs an abstract reachability graph that would witness the satisfiability of CHCs through interpolation.
IC3 \cite{bradley2011sat}  and PDR \cite{vizel2014interpolating} greatly improve the performance of finding the unbounded proof by \textit{incrementally} organized SAT-solving.
It keeps an over-approximation of unsafe states by recursively blocking them and generalizing the blocking lemmas.
Modern solvers embrace the power of SMT-solving to make them extensible to a broader context. 
\textit{Spacer} \cite{komuravelli2013automatic} is a representative work combining SMT-solving techniques with IC3/PDR-styled unbounded model checking, which is also the current default CHC solver in Microsoft Z3 \cite{de2008z3}. It keeps an under-approximation of currently reached states and an over-approximation of unsafe states.
\textit{GSpacer} \cite{vediramana2020global} adds global guidance heuristics 
to enhance Spacer.

\paragraph{Synthesis.}
Methods in this category typically reduce the CHC solving problem into an invariant synthesis problem, which aims to construct an inductive invariant under the \textit{semantic} 
and \textit{syntactic} constraints.
Many of such methods are heavily inspired by the concept of \textit{Counterexample Guided Inductive Synthesis} (CEGIS), where the synthesizer is forced to produce proposals consistent with the counterexamples collected so far.
\textit{Code2Inv} \cite{si2020code2inv, si2018learning} uses a graph neural network to encode the program snippet, and applies reinforcement learning to construct the inductive invariant. 
\textit{FreqHorn} \cite{freq} first generates a grammar that is coherent to the program and a distribution over symbols,
and then synthesizes the invariant using the distribution.
\textit{InvGen} \cite{dillig2013inductive} combines data-driven Boolean learning and synthesis-based feature learning.
\textit{G-CLN} \cite{yao2020learning} uses a template-based differentiable method to generate loop invariants from program traces. 
It can handle some non-linear benchmarks, but manual hyperparameter tuning for each instance is required, which hampers its generality.
In this category, solvers
struggle to scale up to CHCs where multiple invariants exist.


\paragraph{Induction.}
Instead of explicitly constructing the interpretation of unknown predicates, induction-based CHC solvers directly learn such an interpretation by induction. 
The \textit{ICE-learning} framework \cite{garg2014ice} provides a convergent and robust learning paradigm.  HoIce \cite{champion2018hoice} extends the ICE framework to non-linear CHCs.
\textit{ICE-DT} \cite{garg2016learning} introduces a DT algorithm that adapts to a dataset with positive, negative, and implication samples.
\textit{Interval ICE-DT} \cite{xu2020interval} generalizes samples generated from counterexamples based on the UNSAT core technique to closed intervals, which enhances their expressiveness. \textit{LinearArbitrary} \cite{zhu2018data} is the most relevant work to ours. It collects samples by checking and unwinding CHCs, and learns inductive invariants from them using SVM and DT \cite{salzberg1994c4} within the CEGAR-like \cite{cegar} learning iterations. But this approach, due to the ``black-box'' nature, is completely agnostic to the CHC system itself, and much immediate prior knowledge needs to be relearned by the machine learning toolchain.

\subsection{CHCs and Program Safety Verification}
\label{CHC_program_verfication}

An essential application of CHCs is to act as a general format for program safety verification. This generality extends to any imperative programming language and any safety constraints specified in First-Order Logic (FOL). To validate a program, we often use a proof system that generates logical formulas called verification conditions (VCs). Validating VCs implies the correctness of the program. It is also common for VCs to include auxiliary predicates, such as \textit{inductive invariants}. Hence, it is natural that the VCs can be in CHC format, and checking the correctness of the program can be converted to checking the satisfiability of the CHCs. 


The conversion process from program verification to CHC solving is outlined as follows. Consider the program to be verified as a \textit{Control-Flow Graph} (CFG) \cite{bjorner2015horn}. 
A \textit{Basic Blocks} (BBs) in the program is a vertex of the CFG, equivalent to a predicate in the CHC system. Such predicates can be seen as the summary of the BBs, which corresponds to ``\textit{inductive invariants}'' in program verification parlance. 
The edges of the CFG describe the transitions between the BBs. CHCs can encode those transitional relations, in the direction that starts from the BBs of body predicates to the BB of the head predicate.
Then, the solution of the result CHC system corresponds to the correctness of the program.
Such conversion can be fully automated, and numerous tools are developed for different programming languages, such as \textit{SeaHorn} \cite{gurfinkel2015seahorn} and \textit{JayHorn} \cite{kahsai2016jayhorn}.


Table \ref{tab:concept} lists a mapping of program verification concepts to CHC solving concepts.

\begin{table}[h]
\centering
\caption{Mapping of program verification concepts to CHC solving concepts.}
\label{tab:concept}
\renewcommand\arraystretch{1.2} 
\setlength{\tabcolsep}{2mm}{
\begin{tabular}{ll}
\bottomrule
\textbf{Program Verification Concept}  & \textbf{CHC Solving Concept} \\ \bottomrule
Verification Conditions (VCs) & CHC system $\mathcal{H}$         \\ \hline
Initial program state         & Fact      $\mathcal{C}_f$          \\ \hline
Transition                    & Rule $\mathcal{C}_r$                   \\ \hline
\begin{tabular}[c]{@{}l@{}}Assertion\\ Unsafe condition\end{tabular}           & Query $\mathcal{C}_q$             \\ \hline
\begin{tabular}[c]{@{}l@{}}Inductive invariant\\ Loop invariant\end{tabular} & Solution Interpretation $\mathcal{I}^*\left[p\right]$ to predicate $p$  \\ \hline
Counterexample trace          & Refutation proof $\mathcal{R}$  \\ \hline
Reachable program states & Positive samples $s^+$ \\ \hline
Unsafe program states & Negative samples $s^-$ \\
\bottomrule
\end{tabular}}
\end{table}

\subsection{Support Vector Machine}
Support Vector Machines (SVMs) are widely used supervised learning models, especially in small-scale classification problems. In typical SVMs, the quality of a candidate classifier (also called a \textit{hyperplane}) is measured by the margin, defined as its distance from the closest data points, which is often called the support vectors in a vector space. In our work, we only discuss the case that the data are linearly separable, and linear SVMs can be adopted.

In practice, to prevent over-fitting, it is effective to add some slack variables to make the margin ``soft''. The optimizing target of such ``soft-margin'' linear SVM is as follows:

\begin{equation}
\begin{gathered}
\min _{w, \xi_i} \frac{1}{2} w^T w+C \sum_{i=1}^n \xi_i, \\
\text { subject to } y_i\left(w^T \boldsymbol{x}_i+b\right) \geq 1-\xi_i, \xi_i \geq 0, \forall i=1, \ldots, n,
\end{gathered}
\end{equation}
where $w^T \boldsymbol{x}_i+b=0$ is the hyperplane, $\xi_i$ is the i-th slack variable,  $C$ is a hyperparameter that controls the penalty of the error terms. Larger $C$ leads to a larger penalty for wrongly classified samples.

\subsection{Decision Tree}
Decision Tree (DT) is a classic machine learning tool. It provides an explicit, interpretable procedure for classification problems.

The DT algorithms in our tool make a decision based on the \textit{information gain}. We denote the positive and negative samples as $S^+$ and $S^-$, and all samples as $S=S^+\cup S^-$. Then we define the \textit{Shannon Entropy} $\epsilon(S)$:
\begin{equation}
\epsilon(S)=-\frac{\left|S^{+}\right|}{|S|} \log _2 \frac{\left|S^{+}\right|}{|S|}-\frac{\left|S^{-}\right|}{|S|} \log _2 \frac{\left|S^{-}\right|}{|S|},
\end{equation}
and the information gain $\gamma$ with respect to an attribute $f(\boldsymbol{x})\leq c$:

\begin{equation}
\begin{aligned}
\label{infgain}
\gamma(S, f(\boldsymbol{x})\leq c)=\epsilon(S)-&\left(\frac{\left|S_{f(\boldsymbol{x})\leq c}\right| \epsilon(S_{f(\boldsymbol{x})\leq c})}{|S|}+ \right. \\
& \left. \frac{\left|S_{\neg f(\boldsymbol{x})\leq c}\right| \epsilon(S_{\neg f(\boldsymbol{x})\leq c})}{|S|}\right),
\end{aligned}
\end{equation}

where $S_{f(\boldsymbol{x})\leq c}$ and $S_{\neg f(\boldsymbol{x})\leq c}$ are samples satisfying $f(\boldsymbol{x})\leq c$ and samples that do not. The DT prefers to select attributes with higher information gain, i.e., the partition result that has a dominant number of samples in one certain category.

After a DT is learned, by recursively traversing the DT from its root to leaves, we can get a symbolic expression $\mathcal{L}_p$ by taking ``And ($\land$)'' on attributes along a certain path to $\top$ and ``Or ($\lor$)'' between such paths.

\section{Solution Space Analysis}
\label{app:sltana}
In this section, we discuss what properties a solution interpretation for predicate $p$ of a given CHC system $\mathcal{H}$ should hold.  
We use $U_{\mathcal{S}}$ to denote the universe set of any safe zones and positive samples, i.e., $\forall \mathcal{S}_p, \mathcal{S}_p \subseteq  U_{\mathcal{S}}$ and $\forall s^+_p, s^+_p \in  U_{\mathcal{S}}$. $U_{\mathcal{U}}$ is defined similarly as the universe set of any unsafe zones and negative samples.

\begin{lemma}[Disjoint $U_{\mathcal{S}}$ and $U_{\mathcal{U}}$] 
\label{lemma:disj}
For any SAT $\mathcal{H}$, $U_{\mathcal{S}}$ and $U_{\mathcal{U}}$ are disjoint.
\end{lemma}
\begin{proof}
    Suppose in a SAT CHC system $\mathcal{H}$, $U_{\mathcal{S}}$ and $U_{\mathcal{U}}$ is not disjoint, i.e., there is a sample $s$, $s \in U_{\mathcal{S}}$ and $s \in  U_{\mathcal{U}}$. According to the definition of $U_{\mathcal{S}}$ and $U_{\mathcal{U}}$, $s$ is both a positive and a negative sample. According to Lemma \ref{lemma:ssc}, $\mathcal{H}$ is UNSAT, which contradicts the assumption. Therefore, $U_{\mathcal{S}}$ and $U_{\mathcal{U}}$ is disjoint in any SAT CHC system $\mathcal{H}$.\qed
\end{proof}

\begin{figure}
\centering
\includegraphics[width=0.95\textwidth]{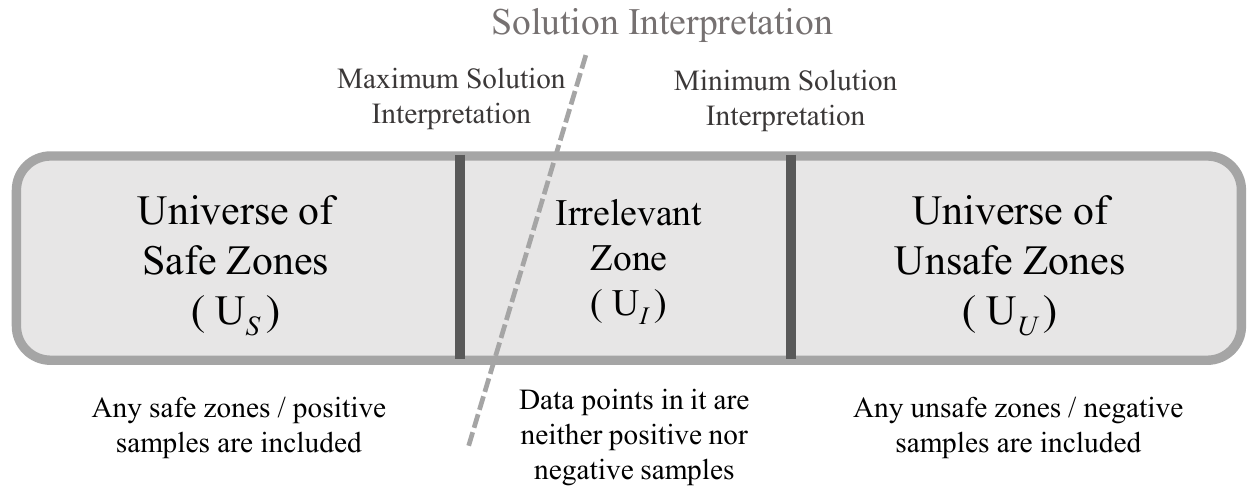}
\caption{
An illustration of the solution space of a CHC system in the state space. The maximum and minimum solution interpretations represent the strongest and weakest solution interpretations respectively. \cite{gu2023optimal} shows the existence of them.
}
\vspace{-0.4cm}
\label{fig:sltspace}
\end{figure}

Given Lemma \ref{lemma:disj}, the state (i.e., sample) space can be divided into three parts shown in Fig. \ref{fig:sltspace}: $U_{\mathcal{S}}$ encapsulating positive samples and safe zones, $U_{\mathcal{U}}$ encompassing negative samples and unsafe zones, and an ``irrelevant zone'' $U_{\mathcal{I}}$ comprising samples that are neither positive nor negative.

\begin{lemma}
\label{lemma:sltinternoits}
A solution interpretation $\mathcal{I}^*\left[p\right]$ cannot intersect with $U_{\mathcal{S}}$ and $U_{\mathcal{U}}$.
\end{lemma}
\begin{proof}
Suppose $\mathcal{I}^*\left[p\right]$ intersects with $U_{\mathcal{S}}$, i.e., $\exists s^+_p, \mathcal{I}^*\left[p\right](s^+_p)=\bot$. This contradicts with Definition \ref{possamp}. 
Suppose $\mathcal{I}^*\left[p\right]$ intersects with $U_{\mathcal{U}}$, i.e., $\exists s^-_p, \mathcal{I}^*\left[p\right](s^-_p)=\top$. This contradicts with Definition \ref{negsamp}. Thus, $\mathcal{I}^*\left[p\right]$ cannot intersect with both $U_{\mathcal{S}}$ and $U_{\mathcal{U}}$. \qed
\end{proof}

We additionally answer the question of what properties a solution interpretation of $\mathcal{H}$ should hold: 

\begin{lemma}[Solution Space]
\label{lemma:sltspace}
$\mathcal{I}$ is a solution interpretation of $\mathcal{H}$ iff: 

1) $\forall p$, $\mathcal{I}\left[p\right]$ does not intersect with $U_{\mathcal{S},p}$ and $U_{\mathcal{U},p}$; 

2) For any $(n+1)$-ary implication sample $\left(s^\rightarrow_{1},\cdots,s^\rightarrow_{n},s^{\rightarrow}_{h}\right)$ of body predicates $(p_1,...,p_n)$ and head predicate $h$, $\mathcal{I}\left[p_1\right]\left(s^\rightarrow_{1}\right)\land\cdots\land \mathcal{I}\left[p_n\right]\left(s^\rightarrow_{n}\right)\rightarrow \mathcal{I}\left[h\right]\left(s^\rightarrow_{h}\right)=\top$.  
\end{lemma}

\paragraph{The Advantage of \name.} Lemma~\ref{lemma:sltinternoits}  and \ref{lemma:sltspace} suggest the importance of finding the coverage of $U_{\mathcal{S}}$ and $U_{\mathcal{U}}$ efficiently. 
If $U_{\mathcal{S}}$ or $U_{\mathcal{U}}$ contains finite samples, in the worst-case scenario, traversing all states of either $U_{\mathcal{S}}$ or $U_{\mathcal{U}}$ is necessary.
Symbolic zonal information, as shown in Definition~\ref{possamp} and \ref{negsamp}, enables efficient computational traversal of a set of states, as learning is not required. 
Even when $U_{\mathcal{S}}$ and $U_{\mathcal{U}}$ contain infinite samples, zones still offer valuable  information by efficiently summarizing the categories of a set of states. This enables the learner to focus on other segments of the state space that may pose challenges for the reasoner to explore.

\section{Implementation Details}
\label{sec:impldet}
Our artifact\footnote{The artifacts are available on this link: \url{https://github.com/Chronosymbolic/Chronosymbolic-Learning}} implementing \name is built in Python 3.10 and utilizes Microsoft Z3 \cite{de2008z3} version 4.8.11.0 as the backend SMT solver. Our tool supports the SMT-LIB2 format and the Datalog (rule-query) format. It is compatible with CHC systems containing single or multiple predicates with arithmetic and Boolean variables.

\subsection{Learner: Data-Driven CHC Solving}
\label{sec:learner_ext}
The \textit{learner} module in our tool leverages an induction-based and CEGAR-inspired \cite{cegar} CHC solving scheme.
It has two sub-modules: the \textit{dataset} and the \textit{machine learning toolchain} (composed of DT and SVM, as shown in Section \ref{mltool}). The dataset stores the positive and negative samples converted from counterexamples\footnote{See Section \ref{sec:converting} for details.} and the corresponding predicates for inductive learning. The machine learning toolchain takes the samples in the dataset as input and outputs a \textit{partition} that can correctly classify all these samples.

\paragraph{SVM.} We apply LIBSVM \cite{CC01a} as our default SVM engine. Positive and negative samples undergo iterative SVM classification until all samples are correctly categorized \cite{zhu2018data}. The resulting hyperplanes are subsequently utilized as attributes within the DT module. Boolean variables are skipped during SVM learning.

\paragraph{DT.}
We offer APIs for C5.0\footnote{https://www.rulequest.com/see5-info.html} \cite{salzberg1994c4}  and CART (Classification and Regression Trees) \cite{breiman1984classification} implemented in scikit-learn \cite{scikit-learn}. In addition to the attributes induced by SVMs, we incorporate several default attributes into DT, making the tool more efficient. These attributes encompass common patterns in inductive invariants such as the octagon abstract domain \cite{mine2006octagon} ($\pm v1 \pm v2\geq c$), 
as well as extensions for specific non-linear operators like \texttt{mod} and \texttt{div}. We also implement an auto-find feature capable of finding the ``\texttt{mod} $k$'' patterns in the CHC system automatically. 
It resembles the idea of some synthesis-based methods, which take often-appearing patterns in the CHC system as a part of the grammar \cite{freq}.
We also provide a visualization tool for analyzing how DTs change through time.

\paragraph{Dataset.} We provide two options for implementation.
\begin{enumerate}
    \item[A.]  \textit{Collecting all data samples until the current iteration and forwarding them all to the learner}. While this approach guarantees accuracy and progress, it can become increasingly burdensome for the learner to induce a partition. 
    \item[B.]  \textit{Maintaining recent $a$ positive and $b$ negative samples in a queue}. This enhances learner efficiency but sacrifices some precision and lacks a progress guarantee. Empirically, both approaches yield nearly the same global performance.
\end{enumerate}

If tentative samples are permitted, they are stored in a distinct dataset, which can also accommodate options A and B. These tentative samples are regularly cleared to maintain (approximate) monotonic improvement. 
In our toolkit, we exclusively employ tentative negative samples and adhere to the methodology outlined in \cite{zhu2018data} to clear these tentative negative samples whenever a positive sample is acquired, ensuring a staged monotonic improvement at the occurrence of positive samples.

\subsection{Reasoner: Zones Discovery}
\label{sec:reasoner_ext}

Our lightweight reasoner implementation aims to enhance hypothesis generation by assisting the learner. It prioritizes simplicity in zones to minimize the computational burden on the backend SMT solver, ensuring smooth execution of the guess-and-check procedure.
\label{ab}
To maintain this simplicity, our implementation follows these principles: 1) \textit{Initialization and Expansion}: The reasoner conducts initialization and expansion primarily at the beginning of the process. 2) \textit{Complexity Control Criteria}: During zone expansion, specific criteria are enforced to prevent zones from becoming excessively complex: Firstly, CHCs with body constraints exceeding a length\footnote{The ``length'' refers to the size of the internal representation of Z3.} of $l_1$ or with free variables exceeding $l_2$ are skipped.
Secondly, a zone reaching a length exceeding $l_3$ triggers the termination of expansion for that zone.
Lastly, a zone introducing ineliminable quantifiers also leads to the termination of the expansion procedure for that zone.

\subsection{Preprocessing of CHC Systems}

To handle CHC systems in various styles and formats, we preprocess the input CHC system in these aspects:
\begin{enumerate}
    \item Rewrite the terms in predicates to ensure that the predicates are only parameterized by distinctive variables (such as $p(v_0,v_1,v_2)$);
    \item Clear and propagate predicates that are $\top$ or $\bot$;
    \item Simplify the inner structure of each CHC if possible.
\end{enumerate}


\section{Experimental Details}
We provide more details in Section \ref{sec:experiment} as follows. The detailed running logs and timing statistics are available in our artifact\footnote{\url{https://github.com/Chronosymbolic/Chronosymbolic-Learning/tree/main/experiment}}.
\subsection{Statistics about the benchmarks}
\label{app:statbench}
Most benchmarks have less than 10 rules, 5 predicates, and 10 variables for each predicate. On average, on all benchmarks under the Chronosymbolic-single setting, we need 1.64 rounds of outer while loop in Alg. \ref{alg}, line 5, and 243.2 iterations of for loop at line 8, so 243.2 counterexamples are generated on average. 

For quantifier elimination (QE), empirically, in most cases, we can do QE when expanding the zones (more than 90\% of cases). However, as described before, considering not slowing down the backend solver, the expansion stops when we cannot do QE.

For non-linear arithmetic benchmarks, we have 19 in total. Aside from them, GSpacer achieves 219/269, and Chronosymbolic-cover achieves 237/269.

In the experiment, $3$ exclusive instances are solved in the Chronosymbolic-single setting (compared with our baselines)\footnote{Exemplar cases are shown at: \url{https://github.com/Chronosymbolic/Chronosymbolic-Learning/tree/main/examples}}.

\subsection{Detailed Settings for Chronosymbolic-cover}
\label{app:detcover}
The Chronosymbolic-cover setting mainly covers the following experiment: 
\begin{enumerate}
    \item[1.] Different expansion strategy of zones (e.g., do not expand at all, small or large limit of size, etc.);
    \item[2.] Different dataset configuration (e.g., whether to enable queue mode on the positive and negative dataset, the length of the queue, how we deal with tentative data samples); 
    \item[3.] Different strategies for scheduling the candidate hypothesis described in Section \ref{tab:cand};
    \item[4.] Different DT settings in Appendix \ref{sec:learner_ext}.
\end{enumerate}

\subsection{Results on Using Different Random Seeds in DT}
\label{app:rnd}
We further examine how randomness affects our system. In this section, we use six different random seeds ($1, 13, 137, 400, 1371, 13711$) in CART for each experiment. We apply dataset option B) in Section \ref{sec:reasoner} for simplicity in analysis. Other configurations remain the same as ``Chronosymbolic-single'' as described in Section \ref{expsetting}. As an example, we only consider the safe instances in our test suite.

\begin{table}[]
\vspace{-0.5cm}
\centering
\caption{Performance evaluation across different random seeds. ``\#solved'' and ``percentage'' stands for solved instances among 235 safe instances. ``avg-time'' is the average time consumed on each instance (including timeout or crashed ones). ``And'' stands for the instances solved under all random seed configurations, while ``Or'' for the instances solved under at least one configuration.}
\label{tab:rnd}
\vspace{0.3cm}
\resizebox{0.8\textwidth}{!}{%
\setlength{\tabcolsep}{5.5mm}{
\begin{tabular}{cccc}
\hline
random seed & \#total & percentage & avg-time (s) \\ \hline
1           & 183     & 77.87\%    & 89.04        \\
13          & 183     & 77.87\%    & 89.89        \\
137         & 177     & 75.32\%    & 94.07        \\
400         & 186     & 79.15\%    & 83.19        \\
1371        & 188     & 80.00\%    & 86.52        \\
13711       & 183     & 77.87\%    & 84.87        \\ \hline
And         & 175     & 74.47\%    & -            \\
Or          & 195     & 82.98\%    & -            \\ \hline
\end{tabular}}
}
\end{table}

As shown in Table \ref{tab:rnd}, the result that our approach consistently solves a substantial number of instances across various random seed configurations demonstrates the robustness of our approach. Different random seeds yield varying partition results for the same dataset, leading to diverse hypotheses exploring distinct directions within the solution space. The ``Or'' result highlights how different seed configurations can collectively cover a broader range of instances.

\subsection{Running Time Analysis}
We provide per-instance time partitioning analysis in our tool. The running time of Z3 (teacher and reasoner), SVM, DT, and reasoner for each instance can be examined through the log \footnote{Check our example: \\\url{https://github.com/Chronosymbolic/Chronosymbolic-Learning/blob/main/experiment/result_safe_summary.log}}. On our main dataset in the ``Chronosymbolic-single'' setting, the average running time for SVM, DT, Z3 are 26.68s, 4.74s, 14.58s, and 1.29s respectively.

\end{document}